\documentclass[prl,aps,twocolumn,showpacs]{revtex4-1}

\usepackage{graphicx}% Include figure files
\usepackage{dcolumn}% Align table columns on decimal point
\usepackage{bm}% bold math
\usepackage{dsfont}
\usepackage{amssymb,amsmath}
\usepackage{sidecap}
\usepackage{wrapfig}

\usepackage{graphicx}
\usepackage{leftidx}
\usepackage{color}

\newcommand{\ket}[1]{\left| #1 \right>} % for Dirac bras
\newcommand{\bra}[1]{\left< #1 \right|} % for Dirac kets
\newcommand{\up}{\uparrow}
\newcommand{\dw}{\downarrow}

\begin{document}

\title{Dynamics of many-body localisation in a translation invariant quantum glass model}

\author{Merlijn van Horssen, Emanuele Levi, Juan P. Garrahan}
\affiliation
{School of Physics and Astronomy, University of Nottingham, Nottingham, NG7 2RD, United Kingdom}

\begin{abstract}
We study the real-time dynamics of a translationally invariant quantum spin chain, based on the East kinetically constrained glass model, in search for evidence of many-body localisation in the absence of disorder.  Numerical simulations indicate a change, controlled by a coupling parameter, from a regime of fast relaxation---corresponding to thermalisation---to a regime of very slow relaxation.  This slowly relaxing regime is characterised by dynamical features usually associated with non-ergodicity and many-body localisation (MBL): memory of initial conditions, logarithmic growth of entanglement entropy, and non-exponential decay of time-correlators.  We show that slow relaxation is a consequence of sensitivity to spatial fluctuations in the initial state.  While numerics indicate that certain relaxation timescales grow markedly with size, our finite size results are consistent both with an MBL transition, expected to only occur in disordered systems, or with a pronounced quasi-MBL crossover.
\end{abstract}

\maketitle

%%%%%%%%%%%%%%%%%%%%%%%%%%%%%%%%%%%%%%%%%%%%%%
%%%%%%%%%%%%%%%%%%%%%%%%%%%%%%%%%%%%%%%%%%%%%%%
%%%%%%%%%%%%%% INTRODUCTION %%%%%%%%%%%%%%%%%%%
%%%%%%%%%%%%%%%%%%%%%%%%%%%%%%%%%%%%%%%%%%%%%%%
%%%%%%%%%%%%%%%%%%%%%%%%%%%%%%%%%%%%%%%%%%%%%%%

\textit{Introduction.}--
The framework of many-body localisation (MBL) \cite{[For a recent review see ]Nandkishore2015} has been shown to be a powerful new approach in the study of non-equilibrium dynamics in closed interacting quantum systems. Since its initial proposal \cite{Basko, *Oganesyan,*Altshuler1997,*Gornyi2005}, abundant \cite{Pal,Berkelbach,*Canovi,*Bardarson,*Serbyn,*Huse2014a,Imbrie,Andraschko2014,*Kondov2015,*Serbyn2014a,*Yao2014a,*Errico2014,*Serbyn2014,Torres-Herrera2015,*Bera2015,*Vasseur2014,*Agarwal2015,*Lev2015,*Laumann2014,*Pekker2014,*Chandran2014} numerical and analytical evidence has accumulated for the existence of MBL, and protocols for realising MBL in particular systems, such as ultracold atomic gases, have been proposed.  MBL has furthermore been observed recently in an experiment on interacting Fermi gases \cite{Bloch}.  While the existence of MBL in systems with explicit quenched disorder is now mostly uncontroversial, it is less clear whether a similar transition can be present in disorder-free transitionally invariant systems \cite{Grover,Schiulaz1,Hickey,Yao,Schiulaz2015,Carleo2012,DeRoeck,Papic2015,Kim2015}.

Aspects of MBL have been observed in particular models without disorder, such as certain mixtures of light and heavy particles \cite{Schiulaz1,Yao}. localisation in such disorder-free systems would occur due to the interplay between interactions and inhomogeneous initial conditions, leading to non-ergodic behaviour where relaxation times would diverge with system size.  In contrast, a general mechanism for delocalisation has been suggested in \cite{DeRoeck}, by which the presence of mobile resonant spots act as carriers of energy or charge. At present it is unclear whether true MBL can occur in a finite-dimensional translation invariant system, and \emph{quasi-}MBL has been suggested \cite{Yao} for cases where transient MBL behaviour is followed by the restoration of ergodicity on long timescales.  (The situation is reminiscent in some sense to that of the classical glass transition problem, where it is debated whether relaxation times diverge at finite temperature with a transition to an ideal glass state, or whether relaxation times are very long but remain finite for all non-zero temperature.  See e.g.\ the discussion in  Ref.\ \cite{Biroli2013}.)

In this work we argue for a transition between an ergodic (delocalised) phase and an MBL phase in a specific translation invariant system.  Generally, a disorderless delocalised-to-MBL phase transition is conjectured to be related to a quantum glass transition \cite{Nandkishore2015}. In particular, in Ref.\ \cite{Hickey} indications of such a delocalised-to-MBL transition were found in a one-dimensional quantum spin chain based on the classical glassy Fredrickson-Anderson model \cite{Fredrickson,Ritort}.  Here we study an even simpler system---and for which MBL-like dynamics is even more marked---a disorderless quantum spin chain based on the (infinite temperature) East glass model \cite{Jackle,Ritort,Faggionato,*Chleboun2013}.  We find that in this model a single coupling parameter controls a significant change in the dynamics, from a regime of fast relaxation and thermalisation, to one where some relaxation times appear to diverge exponentially with system size.  We show that in this slowly relaxing regime the system exhibits many of the dynamical features associated with MBL: logarithmic spreading of excitations, lack of thermalisation, initial state dependence at long times and logarithmic growth of entanglement entropy.  While these observations are highly suggestive of MBL we cannot discard quasi-MBL behaviour as an alternative explanation.

%%%%%%%%%%%%%%%%%%%%%%%%%%%%%%%%%%%%%%%%%%%%%%
%%%%%%%%%%%%%%%%%%%%%%%%%%%%%%%%%%%%%%%%%%%%%%%
%%%%%%%%%%%%%%%%%% MODEL %%%%%%%%%%%%%%%%%%%%%%
%%%%%%%%%%%%%%%%%%%%%%%%%%%%%%%%%%%%%%%%%%%%%%%
%%%%%%%%%%%%%%%%%%%%%%%%%%%%%%%%%%%%%%%%%%%%%%%

\begin{figure}
	\includegraphics[trim = 0mm 0mm 0mm 0mm, clip, width=\columnwidth]{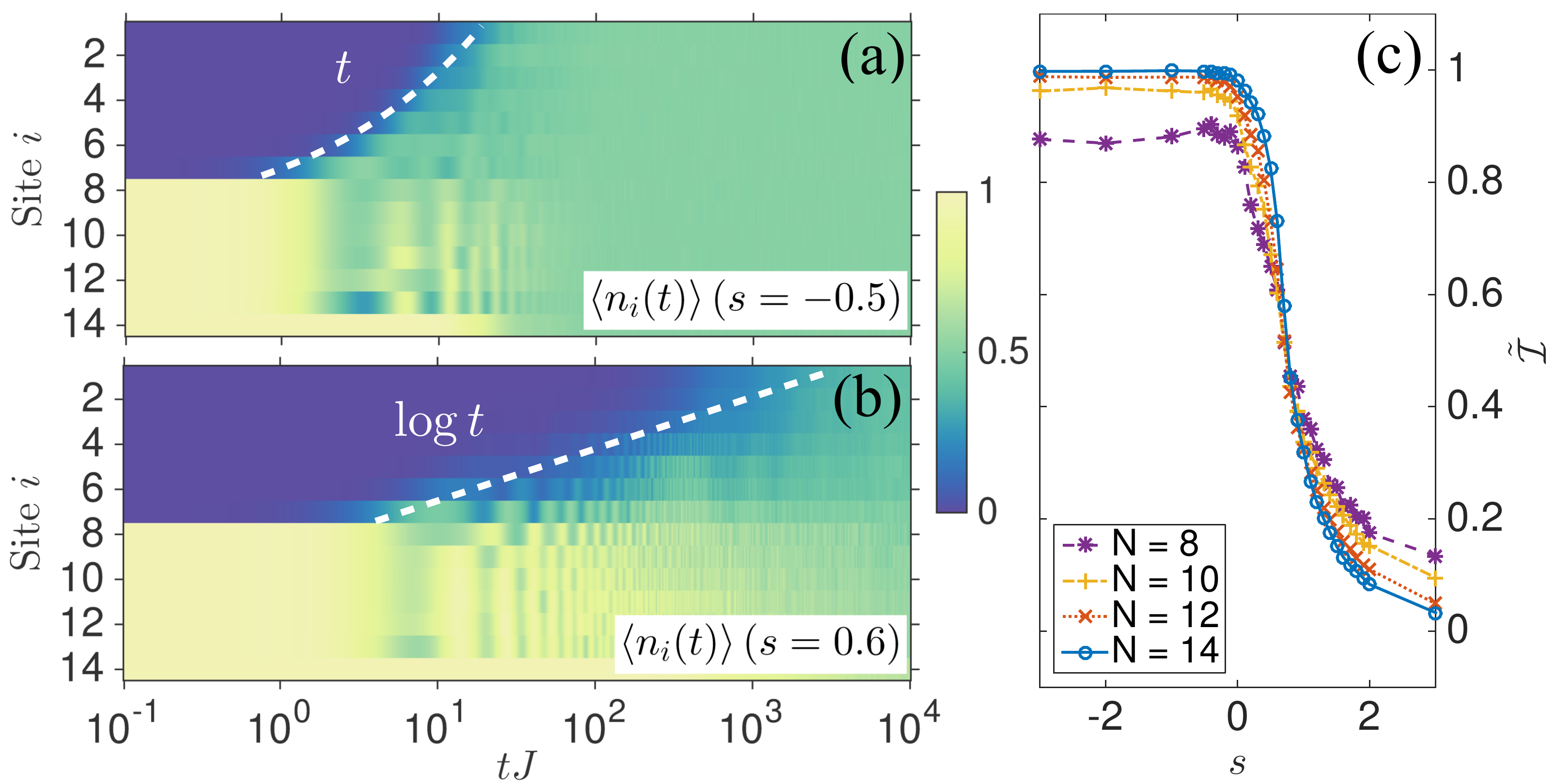}
\caption{(Colour online) Time evolution of site densities $\langle n_{i}(t) \rangle$ starting in the half-filled initial state $\mid\downarrow\cdots\downarrow\uparrow\cdots\uparrow\rangle$ in the thermalising regime, $s = -0.5$ (a), and in the MBL-like regime, $s = 0.6$ (b), showing a linear (ballistic) spreading of the state and a logarithmic spread, respectively. (c): equilibrium value $\overline{\mathcal{I}}_{\mathrm{eq}}$ of the initial participation ratio $\mathcal{I}(t)$, averaged over all half-filling initial states.}
\label{fig:1}
\end{figure}

\textit{Model.}--
We consider a translation invariant spin-$1/2$ chain with Hamiltonian operator
\begin{equation}
 \label{eq:hamiltonian}
 H=\frac{J}{2}\sum_{k=1}^{N}n_k-\frac{J}{2}\sum_{k=1}^{N}e^{-s}\sigma^{x}_k n_{k+1}.
\end{equation}
Here $J$ is an energy scale, which, without loss of generality, we take to be $J=1$.
On the other hand, varying the parameter $s$, which controls the coupling between sites, drastically changes the dynamical behaviour of the model as can be observed in Fig.\ \ref{fig:1}.
This Hamiltonian is closely related to the master operator of the classical East glass model \cite{Jackle,Ritort,Faggionato,*Chleboun2013}.  In fact, if used to generate imaginary time dynamics, the $s=0$ Hamiltonian in Eq.\ \eqref{eq:hamiltonian} is equivalent to the master operator of the classical East model at infinite temperature [i.e., $s=0$ is a Rokhsar-Kivelson point \cite{Rokhsar1988,Castelnovo2005} of \eqref{eq:hamiltonian}].  
At $s \neq 0$ the operator $H$ is a so-called tilted generator \cite{Touchette2009}, whose $s$-dependent ground state is the cumulant generating function of the ``dynamical activity" \cite{Lecomte2007}.  The ground state of the Hamiltonian \eqref{eq:hamiltonian} is known to change discontinuously at $s=0$ \cite{Garrahan2007,*Garrahan2009}. (See also the discussion in \cite{Hickey} for a related model.)

\textit{Classical mechanism for relaxation.}-- In the classical stochastic dynamics generated by \eqref{eq:hamiltonian} when $s=0$, configurations are represented by Fock states in the $z$ basis.  The key feature of these dynamics is that changes of configuration occur via individual spin-flips only on sites which have a right nearest neighbour with $\sigma^z=1$.  This means that a region of space with contiguous down spins can only be relaxed from its rightmost boundary, and as a result, longer regions of this type take longer to relax (as the connecting dynamical pathway is required to show more collective features).  
This ``rare region'' behaviour is what makes the East model glassy at low temperatures (when the ``facilitating'' up spins are sparse) \cite{Jackle,Ritort,Faggionato,*Chleboun2013}.  Furthermore, fluctuations in the initial configuration associated to these down spin domains lead to fluctuations in the spatial relaxation pattern---known as ``dynamical heterogeneity''---of the classical East model \cite{Garrahan2002}.  As we will see below, analogous initial state fluctuations strongly affect the coherent dynamics generated by Eq.\ \eqref{eq:hamiltonian} for general $s$.

\textit{Dynamical features.}--
Considering pure initial states which lack translational invariance reveals dynamical features of our model when unitarily evolving the system with the Hamiltonian $H$. As the system is closed and subject to recurrence one has to consider time--integrated observables to account for relaxation, or limit the dynamics to a small portion of the recurrence cycle \cite{Eisert2014,[See also the recent review ]Gogolin2015}.  

We find two distinct dynamical regimes parametrized by $s$. For $s < 0$ there is fast relaxation and ballistic spreading of initial excitations (up spins), as illustrated in Fig.\ \ref{fig:1}(a). In contrast, the $s>0$ regime exhibits slow relaxation and logarithmic spreading of initial excitations, as shown in Fig.\ \ref{fig:1}(b).  In both figures the initial state is a half-filling Fock state 
(a Fock vector with equal number of up and down spins in the $z$ basis) where the first half of spins are down and the rest is up.  We now argue that the two regimes are distinguished by thermalisation when $s<0$ and lack of thermalisation for $s>0$, and the non-thermalising regime in particular exhibits many of the features of a MBL system.

To probe the dynamics for the possible occurrence of a thermalising-to-MBL transition we consider the dynamical inverse participation ratio (IPR) \cite{Brown2008,*Dukesz2009,*Haake2010} $\mathcal{I}(t) = 2^{-N} \left( \sum_{\mathbf{i}} \vert \langle \psi(t) \vert \mathbf{i} \rangle \mid^{4} \right)^{-1}$ associated to the state $\vert \psi(t) \rangle$ at time $t>0$. 
Fig.\ \ref{fig:1}(c) shows the equilibrium
\footnote{Many-body quantum system are expected to ``equilibrate'' \cite{Gogolin2015}, in the sense of the state becoming close to the time-averaged state, at long enough times under fairly general conditions (such as no energy gap degeneracy).  
This is independent on whether thermalisation is achieved or not (for example this so-called equilibrium state could be initial state dependent).}
value of the IPR $\overline{\mathcal{I}}_{\mathrm{eq}}$, averaged over all half-filling initial states. The quantity $\overline{\mathcal{I}}_{\mathrm{eq}}$ acts as an indicator of localisation of the equilibrium state in the Fock basis; in particular, its value changes from approximately unity for $s<0$ to $\mathcal{O}(N^{-1})$ for $s>0$, suggesting that at the point $s \approx 0$ a MBL transition might occur.

\begin{figure}
	\includegraphics[trim = 0mm 0mm 0mm 0mm, clip, width=\columnwidth]{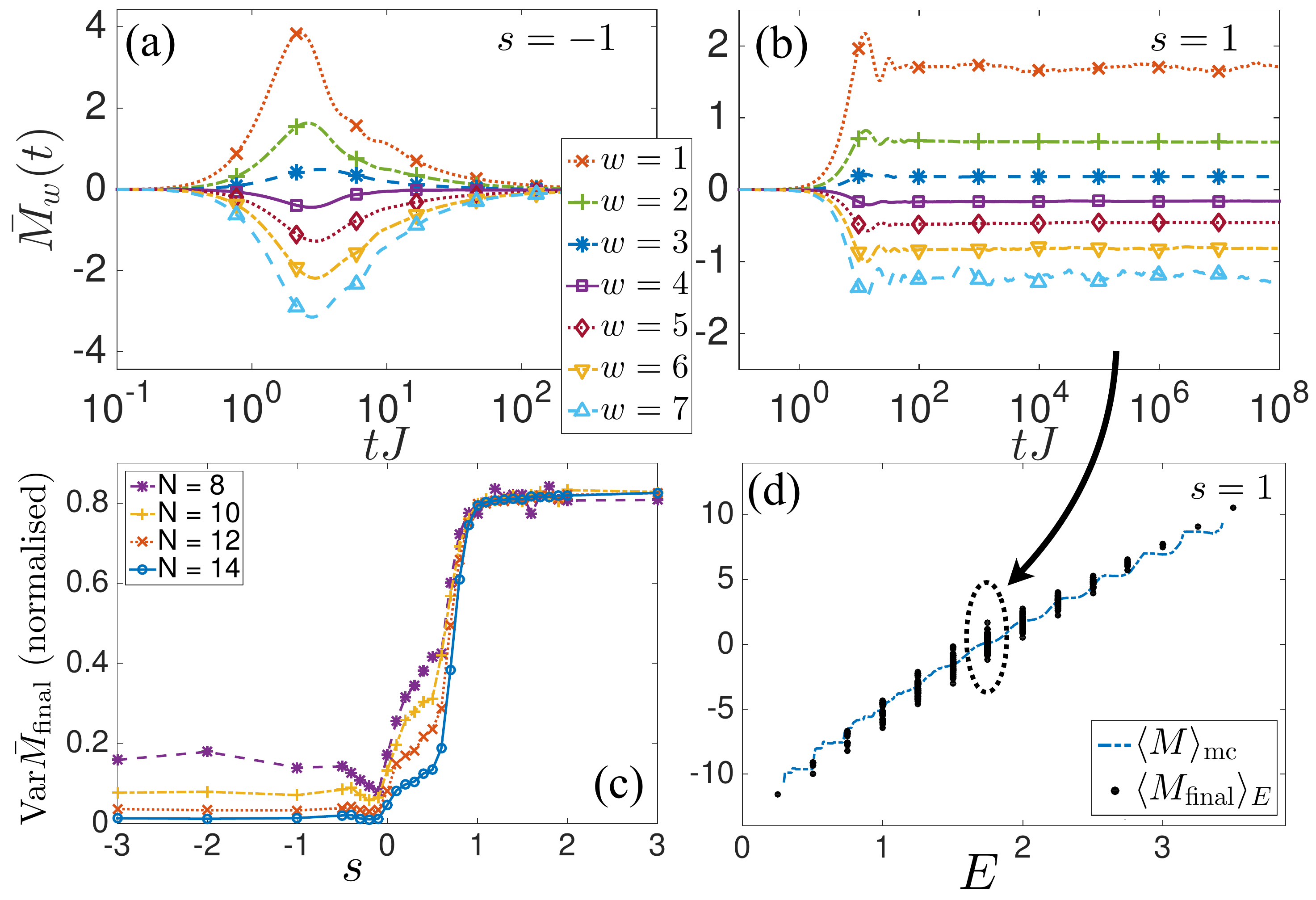}
\caption{(Colour online) (a,b): time-averaged magnetization $\overline{M}_{w}(t)$ resolved by the largest interval $w$ between excitations in the initial state (c): initial state dependence quantified by $\mathrm{var}[M]_{\text{eq}}/\mathrm{var}[M]_{\text{max}}$
(d): comparison between the equilibrium magnetization $\overline{M}_{\mathrm{eq}}$, and its microcanonical average $\langle M \rangle_{\mathrm{mc}}$. The highlighted points are the equilibrium values $\overline{M}_{\mathrm{eq}}$ obtained in panel (b), corresponding to initial states with energy $E \approx 1.7$.
}      
\label{fig:2}
\end{figure}

\textit{Initial state dependence.}--
We consider the dynamical features of the system starting from a pure initial state picked randomly from the half-filling sector.  We characterise the initial state dependence by associating to each initial classical spin configuration a quantity $w$, defined as the maximal number of consecutive down spins, taking into account periodic boundary conditions (e.g. $w=3$ for a Fock vector $\ket{\dw\up\dw\dw\up\up\dw\up\dw\dw}$).

\begin{figure*}
	\includegraphics[trim = 0mm 0mm 0mm 0mm, clip, width=\textwidth]{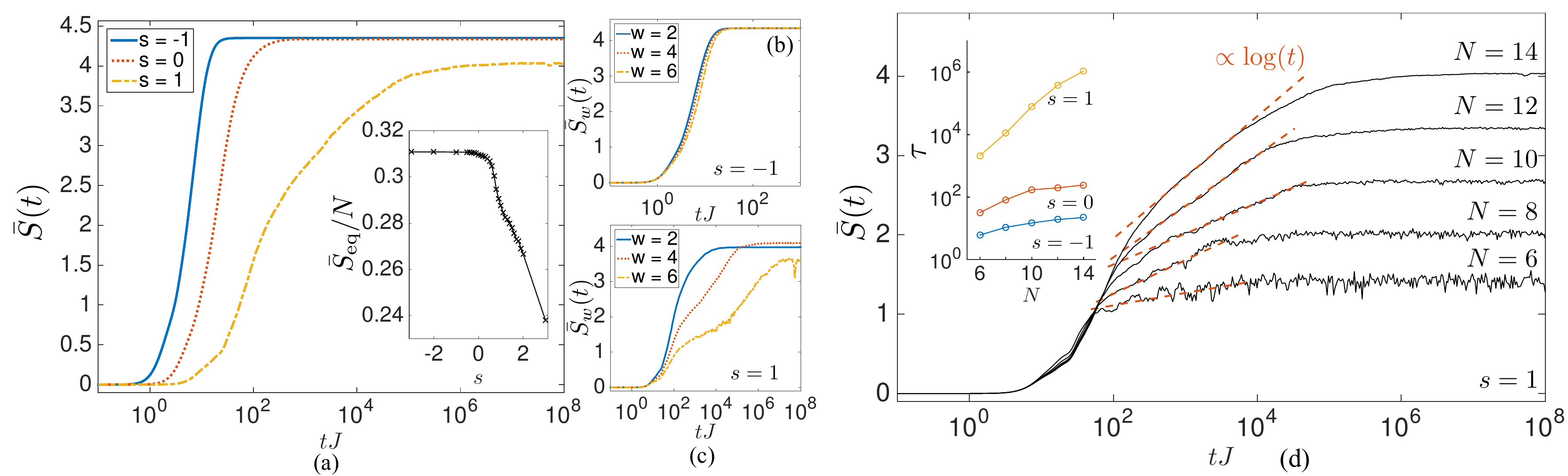}
\caption{(Colour online) (a): entanglement entropy $\overline{S}(t)$ (with respect to an equal bipartition of the spin chain) showing qualitative difference as $s$ is varied. Inset: equilibrium value $\overline{S}_{\text{eq}}$ of entanglement entropy as function of $s$ and $N$, suggesting a transition from the thermalising regime $s<0$ to the MBL-like regime $s > 0$. (c,d): average entanglement entropies $\overline{S}_{w}(t)$ restricted to initial states with domain size $w$, showing that initial state dependence is absent in the thermalising phase.  (d): Characterization of the logarithmic regime in the growth of the entanglement entropy $\overline{S}
(t)$. Dashed lines shows best logarithmic fit  $\overline{S}(t) \propto \log{t}$ for each value of $N$. Inset: estimated relaxation time $\tau$ associated to $\overline{S}(t)$. 
}
\label{fig:3}
\end{figure*}

Long-time dependence of single realizations on the initial state indicates a lack of thermalisation, and is considered a candidate indicator of many-body localisation \cite{Pal,Oganesyan2007,Huse2014a,Gogolin2011}.
We find that the equilibrium state in the regime $s>0$ shows non-thermalising behavior by retaining memory of initial conditions even at long timescales.
This can be seen by considering the time-integrated total magnetization in the $z$-direction $ M(t) = t^{-1} \int_{0}^{t}m(t')dt'$, where $m(t) = \bra{\varphi(t)} \sum_{k} \sigma_{k}^{z} \ket{\varphi(t)}$. 
In Fig.\ \ref{fig:2}(a-b) we display this quantity respectively in the thermalising and non-ergodic regimes resolved according to the maximum distance of excitations present in the initial state Fock vector.  We denote with $\overline{M}_{w}(t)$ the average over realizations restricted to initial states with maximum distance of $w$ sites between excitations.

In the regime $s<0$ the system does not retain memory of the initial state in the long-time limit, as $\overline{M}_{w}(t) \rightarrow 0$ for all values of $w$.
When $s>0$ on the other hand the long-time values of $\overline{M}_{w}(t)$ show a clear dependence on $w$, suggesting an equilibration which depends on the initial conditions, and as such no thermalisation occurs.
This lack of thermalisation is reflected in the time-dependent variance (with respect to initial state realizations) $\mathrm{var}[M(t)]$. In Fig \ref{fig:2}(c) we make the potential transition apparent by plotting the quantity $\mathrm{var}[M]_{\text{eq}}/\mathrm{var}[M]_{\text{max}}$, where $\mathrm{var}[M]_{\text{max}} = \max_{t>0} \mathrm{var}[M(t)]$ and $\mathrm{var}[M]_{\text{eq}}$ denotes the equilibrium value of $\mathrm{var}[M(t)]$.  This scaled variance compares the maximum transient spread of the integrated magnetization (a variation which is expected since at intermediate times different initial states will evolve differently) to the spread at long times 
(a variation which only occurs in the absence of thermalisation).

Further evidence for the absence of thermalisation for $s>1$ is shown in Fig.\ \ref{fig:2}(d) where we plot the equilibrium values of the magnetization $\overline{M}_{\mathrm{eq}}$ along with its microcanonical average expectation value (at infinite temperature), $\langle M \rangle_{\mathrm{mc}}$, as a function of energy \cite{Rigol2008,*Rigol2012}.
Here it can be seen that the microcanonical prediction does not reproduce the long-time average when the initial state energy lies in the bulk of the spectrum.  In this case $\overline{M}_{\mathrm{eq}}$ strongly depends on the initial conditions; these results suggest a breakdown of the eigenstate thermalisation hypothesis leading to absence of thermalisation \cite{Srednicki1994,*Deutsch1991,*Biroli2010}, which in turn is connected to MBL \cite{Pal}.

\textit{Logarithmic growth of entanglement entropy.}--
Slow growth of the entanglement entropy has been associated to a MBL phase in both translation invariant \cite{Yao} and disordered systems \cite{Znidaric2008,Bardarson,*Nanduri2014,*Kjall2014,*Vosk2014,*Friesdorf2014}.  
We consider the entanglement entropy with respect to an equal splitting of the system into two half chains, denoted $A$ and $B$.
Let $S(t)=-\mathrm{Tr}_{A}\rho_{A}(t)\log\rho_{A}(t)$ be the time-dependent entanglement entropy, where $\rho_A(t)=\mathrm{Tr}_B| \psi(t)\rangle \langle\psi(t)|$ is the reduced density matrix on the half chain $A$. We define $\overline{S}(t)$ as the mean entanglement entropy obtained by taking the average over the $S(t)$ associated to all possible initial states in the half filling sector; additionally, to maintain translation invariance, $\overline{S}(t)$ is also averaged over all possible locations of the cut between half chains $A$ and $B$.

In Fig.\ \ref{fig:3}(a) we compare the evolution of the average entropy $\overline{S}(t)$ for different values of $s$, starting from the same initial half-filling ensemble.  Clearly evident is a difference of behavior in the thermalising ($s<0$) and the non-thermalising $(s>0)$ regimes: the former is characterized by an exponential growth which culminates in the equilibrium entropy $\overline{S}_{\text{eq}}$ which is independent of $s$. In the non-thermalising regime on the other hand, after an initial fast growth we observe a slow logarithmic increase which extends over several order of magnitude of $Jt$. The equilibrium value $\overline{S}_{\text{eq}}$ in this regime drops significantly for increasing $s$, as shown in the inset in Fig.\ \ref{fig:3}(a).

The time evolution of $\overline{S}(t)$ can be understood qualitatively as follows:  As previously discussed the typical initial state will contain a largest domain of down spins of length $w$. The initial linear increase of $\overline{S}(t)$ corresponds to the mixing of the state in the regions in away from this empty interval. Once the state is completely mixed over these regions the only increase in entanglement can come from propagating excitations into the $w$ down spin domain,  which as shown in Figs.\ \ref{fig:1}(a,b) is linear in time for $s<0$ and logarithmic for $s>0$.   
In the thermalising regime, $s<0$, there is no difference in the growth of the entanglement entropy starting from states with different $w$, and equilibration is fast.  In contrast, in the non-thermalising regime, $s>0$, this growth shows a pronounced dependence on the value of $w$ in the initial state. 

The range of logarithmic growth of the entropy in the $s>0$ regime increases with system size, as shown in Fig.\ \ref{fig:3}(d).  We can extract a relaxation time $\tau$ as the time the 
$\bar{S}(t)$ reaches $\bar{S}_{\mathrm{eq}}$, for example such that $\left|\bar{S}_{\mathrm{eq}}-\bar{S}(\tau)\right|=\varepsilon$ with $\varepsilon=0.01$ (the precise value of this cutoff is unimportant) \footnote{For small values of $N$ we define $\overline{S}_{\text{eq}}$ as the time average over the final plateau, ensuring that the system does not undergo renewals.}.  The inset to Fig.\ \ref{fig:3}(d) we shows the scaling of the relaxation time $\tau$ for different values of $s$.  For $s>0$, $\tau$ it appears to grow exponentially with the system size in the range of sizes which we were able to consider.  This exponential divergence with size is a hallmark of the loss of ergodicity and of MBL.

\textit{Polarization relaxation.}-- Finally, we consider the relaxation of fluctuations at different wavelengths.  To do this we consider the (infinite temperature) averaged time-correlator of the Fourier transformed longitudinal magnetization, $D(k,t)={\rm Tr}[e^{-iHt} F^{\ast}(k) e^{i H t} F(k)]$, where $F(k) = \sum_{j} \sigma^{z}_{j} e^{i k j}$.  The relaxation of this polarization correlator can be used to quantify the transport of spin perturbations \cite{Pal,Yao} and thus can probe for MBL.  Since in a translation invariant system of finite size, $D(k,t)$ has to decay to zero at large enough times, the interesting property to look for is in the (transient) behaviour of the relaxation times \cite{Yao}. 

In Fig.\ we show $D(k,t)$ as a function of time for all the wave vectors $k$ of a system of size $N=12$ (the largest we were able to simulate for this kind of correlator). For the thermalising case of $s=-1$, Fig.\ 4(a), all correlators decay exponentially and there is no $k$-dependence in their relaxation.  In contrast, for the non-thermalising case of $s=1$, Fig.\ 4(b), relaxation is non-exponential and timescales are $k$-dependent.  Figure 4(c) shows estimates of the relaxation times $\tau$ as a function of $k$ for three system sizes: for $s<0$, $\tau(k)$ is independent of $k$ (and of system size), while for $s>0$, they are strongly $k$ dependent.  In fact, in the non-thermalising phase $\tau(k) \sim e^{1/k \xi}$ for a large range of wavelengths, a similar behaviour as that observed in Ref.\ \cite{Yao}. 

The exponential in wavelength behaviour of the relaxation time can be rationalized as follows.  $D(k,t)$ measures the relaxation over a wavelength $2\pi/k$.  In the MBL-like regime, the initial states that couple to this wavelength and decay the slowest [and thus dominate $D(k,t)$] will be those with $w=2\pi/k$, where $w$ is the size of the largest domain of down spins.  As shown above, such domains decay logarithmically in time, cf.\ Fig.\ 1(b), that is the relaxation into an inactive domain goes as $w \approx \xi \log t$.  Inverting this relation and replacing $w$ by $1/k$ we get the form of $\tau(k)$ above.  In order for this argument to hold these slow initial states have to be numerous in the infinite temperature average.  At the largest wavelength they become rare, and this is be the reason why the exponential scaling is not obeyed in Fig.\ 4(c) for small $k$.  
 
\textit{Conclusions and outlook.}-- In this paper we studied a disorderless quantum spin chain closely related to the classical glassy East model 
to explore a possible MBL transition in a translationally invariant system.  The coherent dynamics of this model displays two regimes; a thermalising regime, and a non-ergodic regime characterised by strong dependence on the initial conditions and breaking of thermalisation.  The non-thermalising regime displays dynamical features characteristic of an MBL phase, in particular a logarithmic spreading of excitations and a logarithmic growth of the entanglement entropy.  The slow relaxation behind these MBL like features can be traced back to spatial fluctuations in the initial state corresponding to rare inactive domains.  While relaxation times seem to diverge with system size for the entanglement entropy, they remain finite (yet anomalous) for dynamic structure factors.  These features could be compatible both with MBL \cite{Nandkishore2015} or with quasi-MBL \cite{Yao}.  In either case, our results suggest a close connection between classical mechanisms for glassy arrest and those responsible for MBL dynamics.

\begin{figure}
	\includegraphics[trim = 0mm 0mm 0mm 0mm, clip, width=\columnwidth]{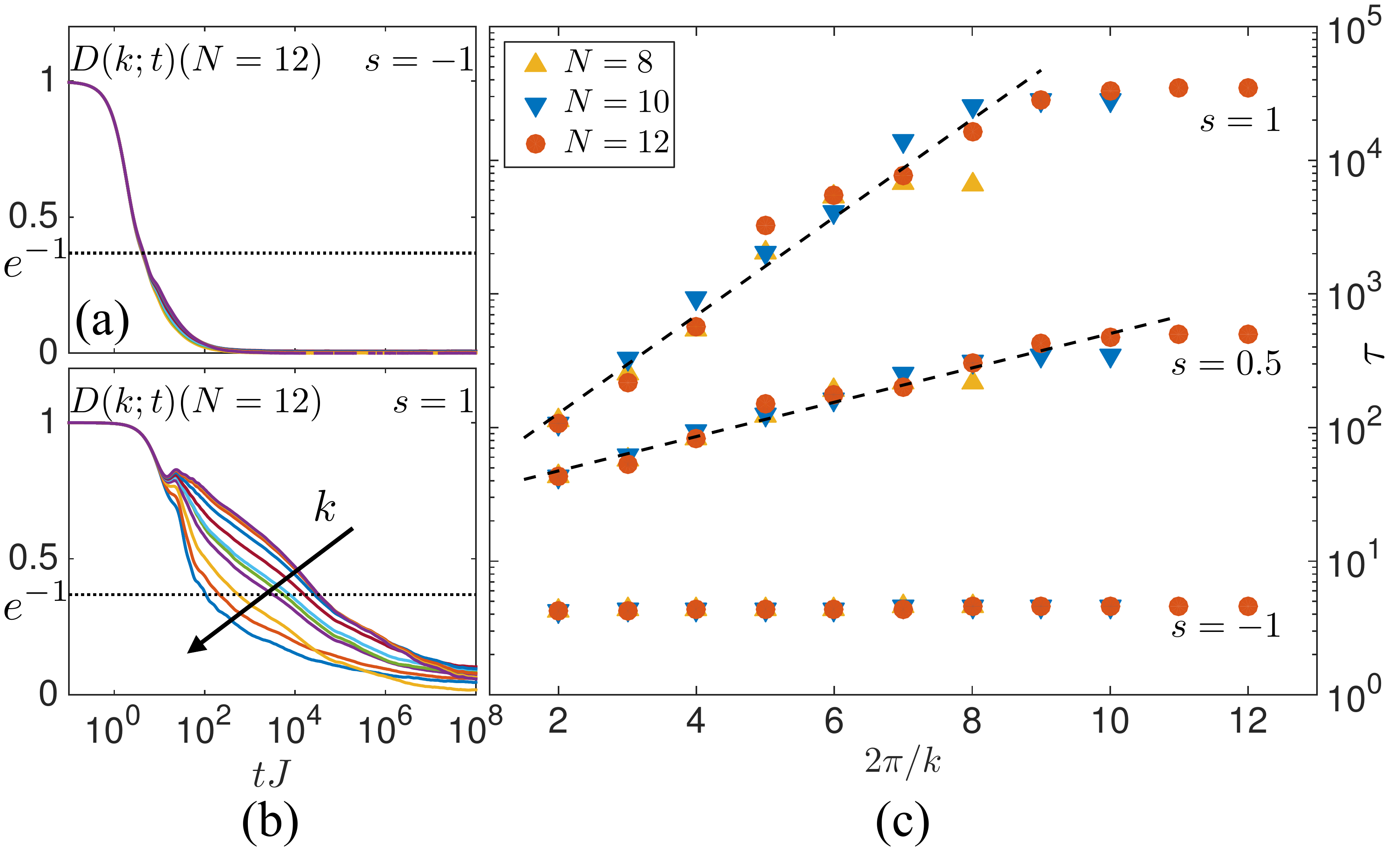}
\caption{(Colour online) Dynamics of the spin polarization $D(k,t)$ for all wave vectors $k$ in a system of size $N=12$. (a): $D(k,t)$ in the thermalising phase, $s=-1$.  (b): $D(k,t)$ in the non-ergodic phase, $s=1$. (c): Wavelength dependent relaxation timescales, estimated by $D(k,\tau)=e^{-1}$, at three values of $s$; the dashed lines indicate $e^{1/k \xi}$ scaling.}
\label{fig:4}
\end{figure}

\textit{Acknowledgments.}-- We thank M. Friesdorf and Z. Papic for feedback and discussions.
This work was supported by EPSRC grant no. EP/L50502X/1 and grant no. EP/J009776/1.


\begin{thebibliography}{68}%
\makeatletter
\providecommand \@ifxundefined [1]{%
 \@ifx{#1\undefined}
}%
\providecommand \@ifnum [1]{%
 \ifnum #1\expandafter \@firstoftwo
 \else \expandafter \@secondoftwo
 \fi
}%
\providecommand \@ifx [1]{%
 \ifx #1\expandafter \@firstoftwo
 \else \expandafter \@secondoftwo
 \fi
}%
\providecommand \natexlab [1]{#1}%
\providecommand \enquote  [1]{``#1''}%
\providecommand \bibnamefont  [1]{#1}%
\providecommand \bibfnamefont [1]{#1}%
\providecommand \citenamefont [1]{#1}%
\providecommand \href@noop [0]{\@secondoftwo}%
\providecommand \href [0]{\begingroup \@sanitize@url \@href}%
\providecommand \@href[1]{\@@startlink{#1}\@@href}%
\providecommand \@@href[1]{\endgroup#1\@@endlink}%
\providecommand \@sanitize@url [0]{\catcode `\\12\catcode `\$12\catcode
  `\&12\catcode `\#12\catcode `\^12\catcode `\_12\catcode `\%12\relax}%
\providecommand \@@startlink[1]{}%
\providecommand \@@endlink[0]{}%
\providecommand \url  [0]{\begingroup\@sanitize@url \@url }%
\providecommand \@url [1]{\endgroup\@href {#1}{\urlprefix }}%
\providecommand \urlprefix  [0]{URL }%
\providecommand \Eprint [0]{\href }%
\providecommand \doibase [0]{http://dx.doi.org/}%
\providecommand \selectlanguage [0]{\@gobble}%
\providecommand \bibinfo  [0]{\@secondoftwo}%
\providecommand \bibfield  [0]{\@secondoftwo}%
\providecommand \translation [1]{[#1]}%
\providecommand \BibitemOpen [0]{}%
\providecommand \bibitemStop [0]{}%
\providecommand \bibitemNoStop [0]{.\EOS\space}%
\providecommand \EOS [0]{\spacefactor3000\relax}%
\providecommand \BibitemShut  [1]{\csname bibitem#1\endcsname}%
\let\auto@bib@innerbib\@empty
%</preamble>
\bibitem [{\citenamefont {Nandkishore}\ and\ \citenamefont
  {Huse}(2015)}]{Nandkishore2015}%
  \BibitemOpen
  \bibfield  {author} {\bibinfo {author} {\bibfnamefont {R.}~\bibnamefont
  {Nandkishore}}\ and\ \bibinfo {author} {\bibfnamefont {D.~A.}\ \bibnamefont
  {Huse}},\ }\href {\doibase 10.1146/annurev-conmatphys-031214-014726}
  {\bibfield  {journal} {\bibinfo  {journal} {Annu. Rev. Condens. Matter
  Phys.}\ }\textbf {\bibinfo {volume} {6}},\ \bibinfo {pages} {15} (\bibinfo
  {year} {2015})}\BibitemShut {NoStop}%
\bibitem [{\citenamefont {Basko}\ \emph {et~al.}(2006)\citenamefont {Basko},
  \citenamefont {Aleiner},\ and\ \citenamefont {Altshuler}}]{Basko}%
  \BibitemOpen
  \bibfield  {author} {\bibinfo {author} {\bibfnamefont {D.}~\bibnamefont
  {Basko}}, \bibinfo {author} {\bibfnamefont {I.}~\bibnamefont {Aleiner}}, \
  and\ \bibinfo {author} {\bibfnamefont {B.}~\bibnamefont {Altshuler}},\ }\href
  {\doibase http://dx.doi.org/10.1016/j.aop.2005.11.014} {\bibfield  {journal}
  {\bibinfo  {journal} {Ann. Phys.}\ }\textbf {\bibinfo {volume} {321}},\
  \bibinfo {pages} {1126 } (\bibinfo {year} {2006})}\BibitemShut {NoStop}%
\bibitem [{\citenamefont {Oganesyan}\ and\ \citenamefont
  {Huse}(2007{\natexlab{a}})}]{Oganesyan}%
  \BibitemOpen
  \bibfield  {author} {\bibinfo {author} {\bibfnamefont {V.}~\bibnamefont
  {Oganesyan}}\ and\ \bibinfo {author} {\bibfnamefont {D.~A.}\ \bibnamefont
  {Huse}},\ }\href {\doibase 10.1103/PhysRevB.75.155111} {\bibfield  {journal}
  {\bibinfo  {journal} {Phys. Rev. B}\ }\textbf {\bibinfo {volume} {75}},\
  \bibinfo {pages} {155111} (\bibinfo {year} {2007}{\natexlab{a}})}\BibitemShut
  {NoStop}%
\bibitem [{\citenamefont {Altshuler}\ \emph {et~al.}(1997)\citenamefont
  {Altshuler}, \citenamefont {Gefen}, \citenamefont {Kamenev},\ and\
  \citenamefont {Levitov}}]{Altshuler1997}%
  \BibitemOpen
  \bibfield  {author} {\bibinfo {author} {\bibfnamefont {B.}~\bibnamefont
  {Altshuler}}, \bibinfo {author} {\bibfnamefont {Y.}~\bibnamefont {Gefen}},
  \bibinfo {author} {\bibfnamefont {A.}~\bibnamefont {Kamenev}}, \ and\
  \bibinfo {author} {\bibfnamefont {L.}~\bibnamefont {Levitov}},\ }\href
  {\doibase 10.1103/PhysRevLett.78.2803} {\bibfield  {journal} {\bibinfo
  {journal} {Phys. Rev. Lett.}\ }\textbf {\bibinfo {volume} {78}},\ \bibinfo
  {pages} {2803} (\bibinfo {year} {1997})}\BibitemShut {NoStop}%
\bibitem [{\citenamefont {Gornyi}\ \emph {et~al.}(2005)\citenamefont {Gornyi},
  \citenamefont {Mirlin},\ and\ \citenamefont {Polyakov}}]{Gornyi2005}%
  \BibitemOpen
  \bibfield  {author} {\bibinfo {author} {\bibfnamefont {I.}~\bibnamefont
  {Gornyi}}, \bibinfo {author} {\bibfnamefont {A.}~\bibnamefont {Mirlin}}, \
  and\ \bibinfo {author} {\bibfnamefont {D.}~\bibnamefont {Polyakov}},\ }\href
  {\doibase 10.1103/PhysRevLett.95.206603} {\bibfield  {journal} {\bibinfo
  {journal} {Phys. Rev. Lett.}\ }\textbf {\bibinfo {volume} {95}},\ \bibinfo
  {pages} {206603} (\bibinfo {year} {2005})}\BibitemShut {NoStop}%
\bibitem [{\citenamefont {Pal}\ and\ \citenamefont {Huse}(2010)}]{Pal}%
  \BibitemOpen
  \bibfield  {author} {\bibinfo {author} {\bibfnamefont {A.}~\bibnamefont
  {Pal}}\ and\ \bibinfo {author} {\bibfnamefont {D.~A.}\ \bibnamefont {Huse}},\
  }\href {\doibase 10.1103/PhysRevB.82.174411} {\bibfield  {journal} {\bibinfo
  {journal} {Phys. Rev. B}\ }\textbf {\bibinfo {volume} {82}},\ \bibinfo
  {pages} {174411} (\bibinfo {year} {2010})}\BibitemShut {NoStop}%
\bibitem [{\citenamefont {Berkelbach}\ and\ \citenamefont
  {Reichman}(2010)}]{Berkelbach}%
  \BibitemOpen
  \bibfield  {author} {\bibinfo {author} {\bibfnamefont {T.~C.}\ \bibnamefont
  {Berkelbach}}\ and\ \bibinfo {author} {\bibfnamefont {D.~R.}\ \bibnamefont
  {Reichman}},\ }\href {\doibase 10.1103/PhysRevB.81.224429} {\bibfield
  {journal} {\bibinfo  {journal} {Phys. Rev. B}\ }\textbf {\bibinfo {volume}
  {81}},\ \bibinfo {pages} {224429} (\bibinfo {year} {2010})}\BibitemShut
  {NoStop}%
\bibitem [{\citenamefont {Canovi}\ \emph {et~al.}(2011)\citenamefont {Canovi},
  \citenamefont {Rossini}, \citenamefont {Fazio}, \citenamefont {Santoro},\
  and\ \citenamefont {Silva}}]{Canovi}%
  \BibitemOpen
  \bibfield  {author} {\bibinfo {author} {\bibfnamefont {E.}~\bibnamefont
  {Canovi}}, \bibinfo {author} {\bibfnamefont {D.}~\bibnamefont {Rossini}},
  \bibinfo {author} {\bibfnamefont {R.}~\bibnamefont {Fazio}}, \bibinfo
  {author} {\bibfnamefont {G.~E.}\ \bibnamefont {Santoro}}, \ and\ \bibinfo
  {author} {\bibfnamefont {A.}~\bibnamefont {Silva}},\ }\href {\doibase
  10.1103/PhysRevB.83.094431} {\bibfield  {journal} {\bibinfo  {journal} {Phys.
  Rev. B}\ }\textbf {\bibinfo {volume} {83}},\ \bibinfo {pages} {094431}
  (\bibinfo {year} {2011})}\BibitemShut {NoStop}%
\bibitem [{\citenamefont {Bardarson}\ \emph {et~al.}(2012)\citenamefont
  {Bardarson}, \citenamefont {Pollmann},\ and\ \citenamefont
  {Moore}}]{Bardarson}%
  \BibitemOpen
  \bibfield  {author} {\bibinfo {author} {\bibfnamefont {J.~H.}\ \bibnamefont
  {Bardarson}}, \bibinfo {author} {\bibfnamefont {F.}~\bibnamefont {Pollmann}},
  \ and\ \bibinfo {author} {\bibfnamefont {J.~E.}\ \bibnamefont {Moore}},\
  }\href {\doibase 10.1103/PhysRevLett.109.017202} {\bibfield  {journal}
  {\bibinfo  {journal} {Phys. Rev. Lett.}\ }\textbf {\bibinfo {volume} {109}},\
  \bibinfo {pages} {017202} (\bibinfo {year} {2012})}\BibitemShut {NoStop}%
\bibitem [{\citenamefont {Serbyn}\ \emph {et~al.}(2013)\citenamefont {Serbyn},
  \citenamefont {Papi\ifmmode~\acute{c}\else \'{c}\fi{}},\ and\ \citenamefont
  {Abanin}}]{Serbyn}%
  \BibitemOpen
  \bibfield  {author} {\bibinfo {author} {\bibfnamefont {M.}~\bibnamefont
  {Serbyn}}, \bibinfo {author} {\bibfnamefont {Z.}~\bibnamefont
  {Papi\ifmmode~\acute{c}\else \'{c}\fi{}}}, \ and\ \bibinfo {author}
  {\bibfnamefont {D.~A.}\ \bibnamefont {Abanin}},\ }\href {\doibase
  10.1103/PhysRevLett.110.260601} {\bibfield  {journal} {\bibinfo  {journal}
  {Phys. Rev. Lett.}\ }\textbf {\bibinfo {volume} {110}},\ \bibinfo {pages}
  {260601} (\bibinfo {year} {2013})}\BibitemShut {NoStop}%
\bibitem [{\citenamefont {Huse}\ \emph {et~al.}(2014)\citenamefont {Huse},
  \citenamefont {Nandkishore},\ and\ \citenamefont {Oganesyan}}]{Huse2014a}%
  \BibitemOpen
  \bibfield  {author} {\bibinfo {author} {\bibfnamefont {D.~A.}\ \bibnamefont
  {Huse}}, \bibinfo {author} {\bibfnamefont {R.}~\bibnamefont {Nandkishore}}, \
  and\ \bibinfo {author} {\bibfnamefont {V.}~\bibnamefont {Oganesyan}},\ }\href
  {\doibase 10.1103/PhysRevB.90.174202} {\bibfield  {journal} {\bibinfo
  {journal} {Phys. Rev. B}\ }\textbf {\bibinfo {volume} {90}},\ \bibinfo
  {pages} {174202} (\bibinfo {year} {2014})}\BibitemShut {NoStop}%
\bibitem [{\citenamefont {Imbrie}()}]{Imbrie}%
  \BibitemOpen
  \bibfield  {author} {\bibinfo {author} {\bibfnamefont {J.~Z.}\ \bibnamefont
  {Imbrie}},\ }\href@noop {} {\ }\Eprint {http://arxiv.org/abs/1403.7837}
  {arXiv:1403.7837} \BibitemShut {NoStop}%
\bibitem [{\citenamefont {Andraschko}\ \emph {et~al.}(2014)\citenamefont
  {Andraschko}, \citenamefont {Enss},\ and\ \citenamefont
  {Sirker}}]{Andraschko2014}%
  \BibitemOpen
  \bibfield  {author} {\bibinfo {author} {\bibfnamefont {F.}~\bibnamefont
  {Andraschko}}, \bibinfo {author} {\bibfnamefont {T.}~\bibnamefont {Enss}}, \
  and\ \bibinfo {author} {\bibfnamefont {J.}~\bibnamefont {Sirker}},\ }\href
  {\doibase 10.1103/PhysRevLett.113.217201} {\bibfield  {journal} {\bibinfo
  {journal} {Phys. Rev. Lett.}\ }\textbf {\bibinfo {volume} {113}},\ \bibinfo
  {pages} {217201} (\bibinfo {year} {2014})}\BibitemShut {NoStop}%
\bibitem [{\citenamefont {Laumann}\ \emph
  {et~al.}(2014{\natexlab{a}})\citenamefont {Laumann}, \citenamefont {Pal},\
  and\ \citenamefont {Scardicchio}}]{Kondov2015}%
  \BibitemOpen
  \bibfield  {author} {\bibinfo {author} {\bibfnamefont {R.}~\bibnamefont
  {Laumann}, \bibfnamefont {C.}}, \bibinfo {author} {\bibfnamefont
  {A.}~\bibnamefont {Pal}}, \ and\ \bibinfo {author} {\bibfnamefont
  {A.}~\bibnamefont {Scardicchio}},\ }\href {\doibase
  10.1103/PhysRevLett.113.200405} {\bibfield  {journal} {\bibinfo  {journal}
  {Phys. Rev. Lett.}\ }\textbf {\bibinfo {volume} {113}},\ \bibinfo {pages}
  {200405} (\bibinfo {year} {2014}{\natexlab{a}})}\BibitemShut {NoStop}%
\bibitem [{\citenamefont {Serbyn}\ \emph
  {et~al.}(2014{\natexlab{a}})\citenamefont {Serbyn}, \citenamefont {Knap},
  \citenamefont {Gopalakrishnan}, \citenamefont {Papi\ifmmode~\acute{c}\else
  \'{c}\fi{}}, \citenamefont {Yao}, \citenamefont {Laumann}, \citenamefont
  {Abanin}, \citenamefont {Lukin},\ and\ \citenamefont {Demler}}]{Serbyn2014a}%
  \BibitemOpen
  \bibfield  {author} {\bibinfo {author} {\bibfnamefont {M.}~\bibnamefont
  {Serbyn}}, \bibinfo {author} {\bibfnamefont {M.}~\bibnamefont {Knap}},
  \bibinfo {author} {\bibfnamefont {S.}~\bibnamefont {Gopalakrishnan}},
  \bibinfo {author} {\bibfnamefont {Z.}~\bibnamefont
  {Papi\ifmmode~\acute{c}\else \'{c}\fi{}}}, \bibinfo {author} {\bibfnamefont
  {Y.}~\bibnamefont {Yao}, \bibfnamefont {N.}}, \bibinfo {author}
  {\bibfnamefont {R.}~\bibnamefont {Laumann}, \bibfnamefont {C.}}, \bibinfo
  {author} {\bibfnamefont {A.}~\bibnamefont {Abanin}, \bibfnamefont {D.}},
  \bibinfo {author} {\bibfnamefont {D.}~\bibnamefont {Lukin}, \bibfnamefont
  {M.}}, \ and\ \bibinfo {author} {\bibfnamefont {A.}~\bibnamefont {Demler},
  \bibfnamefont {E.}},\ }\href {\doibase 10.1103/PhysRevLett.113.147204}
  {\bibfield  {journal} {\bibinfo  {journal} {Phys. Rev. Lett.}\ }\textbf
  {\bibinfo {volume} {113}},\ \bibinfo {pages} {147204} (\bibinfo {year}
  {2014}{\natexlab{a}})}\BibitemShut {NoStop}%
\bibitem [{\citenamefont {Yao}\ \emph {et~al.}(2014)\citenamefont {Yao},
  \citenamefont {Laumann}, \citenamefont {Gopalakrishnan}, \citenamefont
  {Knap}, \citenamefont {M\"uller}, \citenamefont {Demler},\ and\ \citenamefont
  {Lukin}}]{Yao2014a}%
  \BibitemOpen
  \bibfield  {author} {\bibinfo {author} {\bibfnamefont {Y.}~\bibnamefont
  {Yao}, \bibfnamefont {N.}}, \bibinfo {author} {\bibfnamefont
  {R.}~\bibnamefont {Laumann}, \bibfnamefont {C.}}, \bibinfo {author}
  {\bibfnamefont {S.}~\bibnamefont {Gopalakrishnan}}, \bibinfo {author}
  {\bibfnamefont {M.}~\bibnamefont {Knap}}, \bibinfo {author} {\bibfnamefont
  {M.}~\bibnamefont {M\"uller}}, \bibinfo {author} {\bibfnamefont
  {A.}~\bibnamefont {Demler}, \bibfnamefont {E.}}, \ and\ \bibinfo {author}
  {\bibfnamefont {D.}~\bibnamefont {Lukin}, \bibfnamefont {M.}},\ }\href
  {\doibase 10.1103/PhysRevLett.113.243002} {\bibfield  {journal} {\bibinfo
  {journal} {Phys. Rev. Lett.}\ }\textbf {\bibinfo {volume} {113}},\ \bibinfo
  {pages} {243002} (\bibinfo {year} {2014})}\BibitemShut {NoStop}%
\bibitem [{\citenamefont {D'Errico}\ \emph {et~al.}(2014)\citenamefont
  {D'Errico}, \citenamefont {Lucioni}, \citenamefont {Tanzi}, \citenamefont
  {Gori}, \citenamefont {Roux}, \citenamefont {McCulloch}, \citenamefont
  {Giamarchi}, \citenamefont {Inguscio},\ and\ \citenamefont
  {Modugno}}]{Errico2014}%
  \BibitemOpen
  \bibfield  {author} {\bibinfo {author} {\bibfnamefont {C.}~\bibnamefont
  {D'Errico}}, \bibinfo {author} {\bibfnamefont {E.}~\bibnamefont {Lucioni}},
  \bibinfo {author} {\bibfnamefont {L.}~\bibnamefont {Tanzi}}, \bibinfo
  {author} {\bibfnamefont {L.}~\bibnamefont {Gori}}, \bibinfo {author}
  {\bibfnamefont {G.}~\bibnamefont {Roux}}, \bibinfo {author} {\bibfnamefont
  {I.~P.}\ \bibnamefont {McCulloch}}, \bibinfo {author} {\bibfnamefont
  {T.}~\bibnamefont {Giamarchi}}, \bibinfo {author} {\bibfnamefont
  {M.}~\bibnamefont {Inguscio}}, \ and\ \bibinfo {author} {\bibfnamefont
  {G.}~\bibnamefont {Modugno}},\ }\href {\doibase
  10.1103/PhysRevLett.113.095301} {\bibfield  {journal} {\bibinfo  {journal}
  {Phys. Rev. Lett.}\ }\textbf {\bibinfo {volume} {113}},\ \bibinfo {pages}
  {095301} (\bibinfo {year} {2014})}\BibitemShut {NoStop}%
\bibitem [{\citenamefont {Serbyn}\ \emph
  {et~al.}(2014{\natexlab{b}})\citenamefont {Serbyn}, \citenamefont
  {Papi\ifmmode~\acute{c}\else \'{c}\fi{}},\ and\ \citenamefont
  {Abanin}}]{Serbyn2014}%
  \BibitemOpen
  \bibfield  {author} {\bibinfo {author} {\bibfnamefont {M.}~\bibnamefont
  {Serbyn}}, \bibinfo {author} {\bibfnamefont {Z.}~\bibnamefont
  {Papi\ifmmode~\acute{c}\else \'{c}\fi{}}}, \ and\ \bibinfo {author}
  {\bibfnamefont {D.~A.}\ \bibnamefont {Abanin}},\ }\href {\doibase
  10.1103/PhysRevB.90.174302} {\bibfield  {journal} {\bibinfo  {journal} {Phys.
  Rev. B}\ }\textbf {\bibinfo {volume} {90}},\ \bibinfo {pages} {174302}
  (\bibinfo {year} {2014}{\natexlab{b}})}\BibitemShut {NoStop}%
\bibitem [{\citenamefont {Torres-Herrera}\ and\ \citenamefont
  {Santos}(2015)}]{Torres-Herrera2015}%
  \BibitemOpen
  \bibfield  {author} {\bibinfo {author} {\bibfnamefont {E.~J.}\ \bibnamefont
  {Torres-Herrera}}\ and\ \bibinfo {author} {\bibfnamefont {L.~F.}\
  \bibnamefont {Santos}},\ }\href {http://arxiv.org/abs/1501.05662} {\
  (\bibinfo {year} {2015})},\ \Eprint {http://arxiv.org/abs/1501.05662v2}
  {arXiv:1501.05662v2} \BibitemShut {NoStop}%
\bibitem [{\citenamefont {Bera}\ \emph {et~al.}()\citenamefont {Bera},
  \citenamefont {Schomerus}, \citenamefont {Heidrich-Meisner},\ and\
  \citenamefont {Bardarson}}]{Bera2015}%
  \BibitemOpen
  \bibfield  {author} {\bibinfo {author} {\bibfnamefont {S.}~\bibnamefont
  {Bera}}, \bibinfo {author} {\bibfnamefont {H.}~\bibnamefont {Schomerus}},
  \bibinfo {author} {\bibfnamefont {F.}~\bibnamefont {Heidrich-Meisner}}, \
  and\ \bibinfo {author} {\bibfnamefont {J.~H.}\ \bibnamefont {Bardarson}},\
  }\href@noop {} {\ }\BibitemShut {NoStop}%
\bibitem [{\citenamefont {Vasseur}\ \emph {et~al.}(2015)\citenamefont
  {Vasseur}, \citenamefont {Parameswaran},\ and\ \citenamefont
  {Moore}}]{Vasseur2014}%
  \BibitemOpen
  \bibfield  {author} {\bibinfo {author} {\bibfnamefont {R.}~\bibnamefont
  {Vasseur}}, \bibinfo {author} {\bibfnamefont {S.~A.}\ \bibnamefont
  {Parameswaran}}, \ and\ \bibinfo {author} {\bibfnamefont {J.~E.}\
  \bibnamefont {Moore}},\ }\href {\doibase 10.1103/PhysRevB.91.140202}
  {\bibfield  {journal} {\bibinfo  {journal} {Phys. Rev. B}\ }\textbf {\bibinfo
  {volume} {91}},\ \bibinfo {pages} {140202} (\bibinfo {year}
  {2015})}\BibitemShut {NoStop}%
\bibitem [{\citenamefont {Agarwal}\ \emph {et~al.}(2015)\citenamefont
  {Agarwal}, \citenamefont {Gopalakrishnan}, \citenamefont {Knap},
  \citenamefont {M\"uller},\ and\ \citenamefont {Demler}}]{Agarwal2015}%
  \BibitemOpen
  \bibfield  {author} {\bibinfo {author} {\bibfnamefont {K.}~\bibnamefont
  {Agarwal}}, \bibinfo {author} {\bibfnamefont {S.}~\bibnamefont
  {Gopalakrishnan}}, \bibinfo {author} {\bibfnamefont {M.}~\bibnamefont
  {Knap}}, \bibinfo {author} {\bibfnamefont {M.}~\bibnamefont {M\"uller}}, \
  and\ \bibinfo {author} {\bibfnamefont {E.}~\bibnamefont {Demler}},\ }\href
  {\doibase 10.1103/PhysRevLett.114.160401} {\bibfield  {journal} {\bibinfo
  {journal} {Phys. Rev. Lett.}\ }\textbf {\bibinfo {volume} {114}},\ \bibinfo
  {pages} {160401} (\bibinfo {year} {2015})}\BibitemShut {NoStop}%
\bibitem [{\citenamefont {Bar~Lev}\ \emph {et~al.}(2015)\citenamefont
  {Bar~Lev}, \citenamefont {Cohen},\ and\ \citenamefont {Reichman}}]{Lev2015}%
  \BibitemOpen
  \bibfield  {author} {\bibinfo {author} {\bibfnamefont {Y.}~\bibnamefont
  {Bar~Lev}}, \bibinfo {author} {\bibfnamefont {G.}~\bibnamefont {Cohen}}, \
  and\ \bibinfo {author} {\bibfnamefont {D.~R.}\ \bibnamefont {Reichman}},\
  }\href {\doibase 10.1103/PhysRevLett.114.100601} {\bibfield  {journal}
  {\bibinfo  {journal} {Phys. Rev. Lett.}\ }\textbf {\bibinfo {volume} {114}},\
  \bibinfo {pages} {100601} (\bibinfo {year} {2015})}\BibitemShut {NoStop}%
\bibitem [{\citenamefont {Laumann}\ \emph
  {et~al.}(2014{\natexlab{b}})\citenamefont {Laumann}, \citenamefont {Pal},\
  and\ \citenamefont {Scardicchio}}]{Laumann2014}%
  \BibitemOpen
  \bibfield  {author} {\bibinfo {author} {\bibfnamefont {R.}~\bibnamefont
  {Laumann}, \bibfnamefont {C.}}, \bibinfo {author} {\bibfnamefont
  {A.}~\bibnamefont {Pal}}, \ and\ \bibinfo {author} {\bibfnamefont
  {A.}~\bibnamefont {Scardicchio}},\ }\href {\doibase
  10.1103/PhysRevLett.113.200405} {\bibfield  {journal} {\bibinfo  {journal}
  {Phys. Rev. Lett.}\ }\textbf {\bibinfo {volume} {113}},\ \bibinfo {pages}
  {200405} (\bibinfo {year} {2014}{\natexlab{b}})}\BibitemShut {NoStop}%
\bibitem [{\citenamefont {Pekker}\ \emph {et~al.}(2014)\citenamefont {Pekker},
  \citenamefont {Refael}, \citenamefont {Altman}, \citenamefont {Demler},\ and\
  \citenamefont {Oganesyan}}]{Pekker2014}%
  \BibitemOpen
  \bibfield  {author} {\bibinfo {author} {\bibfnamefont {D.}~\bibnamefont
  {Pekker}}, \bibinfo {author} {\bibfnamefont {G.}~\bibnamefont {Refael}},
  \bibinfo {author} {\bibfnamefont {E.}~\bibnamefont {Altman}}, \bibinfo
  {author} {\bibfnamefont {E.}~\bibnamefont {Demler}}, \ and\ \bibinfo {author}
  {\bibfnamefont {V.}~\bibnamefont {Oganesyan}},\ }\href {\doibase
  10.1103/PhysRevX.4.011052} {\bibfield  {journal} {\bibinfo  {journal} {Phys.
  Rev. X}\ }\textbf {\bibinfo {volume} {4}},\ \bibinfo {pages} {011052}
  (\bibinfo {year} {2014})}\BibitemShut {NoStop}%
\bibitem [{\citenamefont {Chandran}\ \emph {et~al.}(2015)\citenamefont
  {Chandran}, \citenamefont {Kim}, \citenamefont {Vidal},\ and\ \citenamefont
  {Abanin}}]{Chandran2014}%
  \BibitemOpen
  \bibfield  {author} {\bibinfo {author} {\bibfnamefont {A.}~\bibnamefont
  {Chandran}}, \bibinfo {author} {\bibfnamefont {I.~H.}\ \bibnamefont {Kim}},
  \bibinfo {author} {\bibfnamefont {G.}~\bibnamefont {Vidal}}, \ and\ \bibinfo
  {author} {\bibfnamefont {D.~A.}\ \bibnamefont {Abanin}},\ }\href {\doibase
  10.1103/PhysRevB.91.085425} {\bibfield  {journal} {\bibinfo  {journal} {Phys.
  Rev. B}\ }\textbf {\bibinfo {volume} {91}},\ \bibinfo {pages} {085425}
  (\bibinfo {year} {2015})}\BibitemShut {NoStop}%
\bibitem [{\citenamefont {{Schreiber}}\ \emph {et~al.}()\citenamefont
  {{Schreiber}}, \citenamefont {{Hodgman}}, \citenamefont {{Bordia}},
  \citenamefont {{L{\"u}schen}}, \citenamefont {{Fischer}}, \citenamefont
  {{Vosk}}, \citenamefont {{Altman}}, \citenamefont {{Schneider}},\ and\
  \citenamefont {{Bloch}}}]{Bloch}%
  \BibitemOpen
  \bibfield  {author} {\bibinfo {author} {\bibfnamefont {M.}~\bibnamefont
  {{Schreiber}}}, \bibinfo {author} {\bibfnamefont {S.~S.}\ \bibnamefont
  {{Hodgman}}}, \bibinfo {author} {\bibfnamefont {P.}~\bibnamefont {{Bordia}}},
  \bibinfo {author} {\bibfnamefont {H.~P.}\ \bibnamefont {{L{\"u}schen}}},
  \bibinfo {author} {\bibfnamefont {M.~H.}\ \bibnamefont {{Fischer}}}, \bibinfo
  {author} {\bibfnamefont {R.}~\bibnamefont {{Vosk}}}, \bibinfo {author}
  {\bibfnamefont {E.}~\bibnamefont {{Altman}}}, \bibinfo {author}
  {\bibfnamefont {U.}~\bibnamefont {{Schneider}}}, \ and\ \bibinfo {author}
  {\bibfnamefont {I.}~\bibnamefont {{Bloch}}},\ }\href@noop {} {\ }\Eprint
  {http://arxiv.org/abs/1501.05661} {arXiv:1501.05661} \BibitemShut {NoStop}%
\bibitem [{\citenamefont {Grover}\ and\ \citenamefont {Fisher}(2014)}]{Grover}%
  \BibitemOpen
  \bibfield  {author} {\bibinfo {author} {\bibfnamefont {T.}~\bibnamefont
  {Grover}}\ and\ \bibinfo {author} {\bibfnamefont {M.~P.~A.}\ \bibnamefont
  {Fisher}},\ }\href {http://stacks.iop.org/1742-5468/2014/i=10/a=P10010}
  {\bibfield  {journal} {\bibinfo  {journal} {J. Stat. Mech.}\ }\textbf
  {\bibinfo {volume} {2014}},\ \bibinfo {pages} {P10010} (\bibinfo {year}
  {2014})}\BibitemShut {NoStop}%
\bibitem [{\citenamefont {{Schiulaz}}\ and\ \citenamefont
  {{M{\"u}ller}}(2014)}]{Schiulaz1}%
  \BibitemOpen
  \bibfield  {author} {\bibinfo {author} {\bibfnamefont {M.}~\bibnamefont
  {{Schiulaz}}}\ and\ \bibinfo {author} {\bibfnamefont {M.}~\bibnamefont
  {{M{\"u}ller}}},\ }in\ \href {\doibase 10.1063/1.4893505} {\emph {\bibinfo
  {booktitle} {American Institute of Physics Conference Series}}},\ \bibinfo
  {series} {American Institute of Physics Conference Series}, Vol.\ \bibinfo
  {volume} {1610}\ (\bibinfo {year} {2014})\ pp.\ \bibinfo {pages}
  {11--23}\BibitemShut {NoStop}%
\bibitem [{\citenamefont {{Hickey}}\ \emph {et~al.}()\citenamefont {{Hickey}},
  \citenamefont {{Genway}},\ and\ \citenamefont {{Garrahan}}}]{Hickey}%
  \BibitemOpen
  \bibfield  {author} {\bibinfo {author} {\bibfnamefont {J.~M.}\ \bibnamefont
  {{Hickey}}}, \bibinfo {author} {\bibfnamefont {S.}~\bibnamefont {{Genway}}},
  \ and\ \bibinfo {author} {\bibfnamefont {J.~P.}\ \bibnamefont {{Garrahan}}},\
  }\href@noop {} {\ }\Eprint {http://arxiv.org/abs/1405.5780} {arXiv:1405.5780}
  \BibitemShut {NoStop}%
\bibitem [{\citenamefont {{Yao}}\ \emph {et~al.}()\citenamefont {{Yao}},
  \citenamefont {{Laumann}}, \citenamefont {{Cirac}}, \citenamefont {{Lukin}},\
  and\ \citenamefont {{Moore}}}]{Yao}%
  \BibitemOpen
  \bibfield  {author} {\bibinfo {author} {\bibfnamefont {N.~Y.}\ \bibnamefont
  {{Yao}}}, \bibinfo {author} {\bibfnamefont {C.~R.}\ \bibnamefont
  {{Laumann}}}, \bibinfo {author} {\bibfnamefont {J.~I.}\ \bibnamefont
  {{Cirac}}}, \bibinfo {author} {\bibfnamefont {M.~D.}\ \bibnamefont
  {{Lukin}}}, \ and\ \bibinfo {author} {\bibfnamefont {J.~E.}\ \bibnamefont
  {{Moore}}},\ }\href@noop {} {\ }\Eprint {http://arxiv.org/abs/1410.7407}
  {arXiv:1410.7407} \BibitemShut {NoStop}%
\bibitem [{\citenamefont {Schiulaz}\ \emph {et~al.}(2015)\citenamefont
  {Schiulaz}, \citenamefont {Silva},\ and\ \citenamefont
  {M\"uller}}]{Schiulaz2015}%
  \BibitemOpen
  \bibfield  {author} {\bibinfo {author} {\bibfnamefont {M.}~\bibnamefont
  {Schiulaz}}, \bibinfo {author} {\bibfnamefont {A.}~\bibnamefont {Silva}}, \
  and\ \bibinfo {author} {\bibfnamefont {M.}~\bibnamefont {M\"uller}},\ }\href
  {\doibase 10.1103/PhysRevB.91.184202} {\bibfield  {journal} {\bibinfo
  {journal} {Phys. Rev. B}\ }\textbf {\bibinfo {volume} {91}},\ \bibinfo
  {pages} {184202} (\bibinfo {year} {2015})}\BibitemShut {NoStop}%
\bibitem [{\citenamefont {Carleo}\ \emph {et~al.}(2012)\citenamefont {Carleo},
  \citenamefont {Becca}, \citenamefont {Schir\'{o}},\ and\ \citenamefont
  {Fabrizio}}]{Carleo2012}%
  \BibitemOpen
  \bibfield  {author} {\bibinfo {author} {\bibfnamefont {G.}~\bibnamefont
  {Carleo}}, \bibinfo {author} {\bibfnamefont {F.}~\bibnamefont {Becca}},
  \bibinfo {author} {\bibfnamefont {M.}~\bibnamefont {Schir\'{o}}}, \ and\
  \bibinfo {author} {\bibfnamefont {M.}~\bibnamefont {Fabrizio}},\ }\href
  {\doibase 10.1038/srep00243} {\bibfield  {journal} {\bibinfo  {journal}
  {Scientific reports}\ }\textbf {\bibinfo {volume} {2}},\ \bibinfo {pages}
  {243} (\bibinfo {year} {2012})}\BibitemShut {NoStop}%
\bibitem [{\citenamefont {De~Roeck}\ and\ \citenamefont
  {Huveneers}(2014)}]{DeRoeck}%
  \BibitemOpen
  \bibfield  {author} {\bibinfo {author} {\bibfnamefont {W.}~\bibnamefont
  {De~Roeck}}\ and\ \bibinfo {author} {\bibfnamefont {F.~m.~c.}\ \bibnamefont
  {Huveneers}},\ }\href {\doibase 10.1103/PhysRevB.90.165137} {\bibfield
  {journal} {\bibinfo  {journal} {Phys. Rev. B}\ }\textbf {\bibinfo {volume}
  {90}},\ \bibinfo {pages} {165137} (\bibinfo {year} {2014})}\BibitemShut
  {NoStop}%
\bibitem [{\citenamefont {Papic}\ \emph {et~al.}()\citenamefont {Papic},
  \citenamefont {Stoudenmire},\ and\ \citenamefont {Abanin}}]{Papic2015}%
  \BibitemOpen
  \bibfield  {author} {\bibinfo {author} {\bibfnamefont {Z.}~\bibnamefont
  {Papic}}, \bibinfo {author} {\bibfnamefont {E.~M.}\ \bibnamefont
  {Stoudenmire}}, \ and\ \bibinfo {author} {\bibfnamefont {D.~A.}\ \bibnamefont
  {Abanin}},\ }\href@noop {} {\ }\Eprint {http://arxiv.org/abs/1501.00477}
  {arXiv:1501.00477} \BibitemShut {NoStop}%
\bibitem [{\citenamefont {Kim}\ and\ \citenamefont {Haah}()}]{Kim2015}%
  \BibitemOpen
  \bibfield  {author} {\bibinfo {author} {\bibfnamefont {I.~H.}\ \bibnamefont
  {Kim}}\ and\ \bibinfo {author} {\bibfnamefont {J.}~\bibnamefont {Haah}},\
  }\href@noop {} {\ }\Eprint {http://arxiv.org/abs/1505.01480}
  {arXiv:1505.01480} \BibitemShut {NoStop}%
\bibitem [{\citenamefont {Biroli}\ and\ \citenamefont
  {Garrahan}(2013)}]{Biroli2013}%
  \BibitemOpen
  \bibfield  {author} {\bibinfo {author} {\bibfnamefont {G.}~\bibnamefont
  {Biroli}}\ and\ \bibinfo {author} {\bibfnamefont {J.~P.}\ \bibnamefont
  {Garrahan}},\ }\href {\doibase 10.1063/1.4795539} {\bibfield  {journal}
  {\bibinfo  {journal} {J. Chem. Phys.}\ }\textbf {\bibinfo {volume} {138}},\
  \bibinfo {eid} {12A301} (\bibinfo {year} {2013})}\BibitemShut {NoStop}%
\bibitem [{\citenamefont {Fredrickson}\ and\ \citenamefont
  {Andersen}(1984)}]{Fredrickson}%
  \BibitemOpen
  \bibfield  {author} {\bibinfo {author} {\bibfnamefont {G.~H.}\ \bibnamefont
  {Fredrickson}}\ and\ \bibinfo {author} {\bibfnamefont {H.~C.}\ \bibnamefont
  {Andersen}},\ }\href {\doibase 10.1103/PhysRevLett.53.1244} {\bibfield
  {journal} {\bibinfo  {journal} {Phys. Rev. Lett.}\ }\textbf {\bibinfo
  {volume} {53}},\ \bibinfo {pages} {1244} (\bibinfo {year}
  {1984})}\BibitemShut {NoStop}%
\bibitem [{\citenamefont {Ritort}\ and\ \citenamefont
  {Sollich}(2003)}]{Ritort}%
  \BibitemOpen
  \bibfield  {author} {\bibinfo {author} {\bibfnamefont {F.}~\bibnamefont
  {Ritort}}\ and\ \bibinfo {author} {\bibfnamefont {P.}~\bibnamefont
  {Sollich}},\ }\href {\doibase 10.1080/0001873031000093582} {\bibfield
  {journal} {\bibinfo  {journal} {Adv. Phys.}\ }\textbf {\bibinfo {volume}
  {52}},\ \bibinfo {pages} {219} (\bibinfo {year} {2003})}\BibitemShut
  {NoStop}%
\bibitem [{\citenamefont {J\"{a}ckle}\ and\ \citenamefont
  {Eisinger}(1991)}]{Jackle}%
  \BibitemOpen
  \bibfield  {author} {\bibinfo {author} {\bibfnamefont {J.}~\bibnamefont
  {J\"{a}ckle}}\ and\ \bibinfo {author} {\bibfnamefont {S.}~\bibnamefont
  {Eisinger}},\ }\href {\doibase 10.1007/BF01453764} {\bibfield  {journal}
  {\bibinfo  {journal} {Z. fur Phys. B}\ }\textbf {\bibinfo {volume} {84}},\
  \bibinfo {pages} {115} (\bibinfo {year} {1991})}\BibitemShut {NoStop}%
\bibitem [{\citenamefont {{Faggionato}}\ \emph {et~al.}()\citenamefont
  {{Faggionato}}, \citenamefont {{Martinelli}}, \citenamefont {{Roberto}},\
  and\ \citenamefont {{Toninelli}}}]{Faggionato}%
  \BibitemOpen
  \bibfield  {author} {\bibinfo {author} {\bibfnamefont {A.}~\bibnamefont
  {{Faggionato}}}, \bibinfo {author} {\bibfnamefont {F.}~\bibnamefont
  {{Martinelli}}}, \bibinfo {author} {\bibfnamefont {C.}~\bibnamefont
  {{Roberto}}}, \ and\ \bibinfo {author} {\bibfnamefont {C.}~\bibnamefont
  {{Toninelli}}},\ }\href@noop {} {\ }\Eprint {http://arxiv.org/abs/1205.1607}
  {arXiv:1205.1607} \BibitemShut {NoStop}%
\bibitem [{\citenamefont {Chleboun}\ \emph {et~al.}(2013)\citenamefont
  {Chleboun}, \citenamefont {Faggionato},\ and\ \citenamefont
  {Martinelli}}]{Chleboun2013}%
  \BibitemOpen
  \bibfield  {author} {\bibinfo {author} {\bibfnamefont {P.}~\bibnamefont
  {Chleboun}}, \bibinfo {author} {\bibfnamefont {A.}~\bibnamefont
  {Faggionato}}, \ and\ \bibinfo {author} {\bibfnamefont {F.}~\bibnamefont
  {Martinelli}},\ }\href@noop {} {\bibfield  {journal} {\bibinfo  {journal} {J.
  Stat. Mech.}\ ,\ \bibinfo {pages} {L04001}} (\bibinfo {year}
  {2013})}\BibitemShut {NoStop}%
\bibitem [{\citenamefont {Rokhsar}\ and\ \citenamefont
  {Kivelson}(1988)}]{Rokhsar1988}%
  \BibitemOpen
  \bibfield  {author} {\bibinfo {author} {\bibfnamefont {D.~S.}\ \bibnamefont
  {Rokhsar}}\ and\ \bibinfo {author} {\bibfnamefont {S.~A.}\ \bibnamefont
  {Kivelson}},\ }\href@noop {} {\bibfield  {journal} {\bibinfo  {journal}
  {Phys. Rev. Lett.}\ }\textbf {\bibinfo {volume} {61}},\ \bibinfo {pages}
  {2376} (\bibinfo {year} {1988})}\BibitemShut {NoStop}%
\bibitem [{\citenamefont {Castelnovo}\ \emph {et~al.}(2005)\citenamefont
  {Castelnovo}, \citenamefont {Chamon}, \citenamefont {Mudry},\ and\
  \citenamefont {Pujol}}]{Castelnovo2005}%
  \BibitemOpen
  \bibfield  {author} {\bibinfo {author} {\bibfnamefont {C.}~\bibnamefont
  {Castelnovo}}, \bibinfo {author} {\bibfnamefont {C.}~\bibnamefont {Chamon}},
  \bibinfo {author} {\bibfnamefont {C.}~\bibnamefont {Mudry}}, \ and\ \bibinfo
  {author} {\bibfnamefont {P.}~\bibnamefont {Pujol}},\ }\href@noop {}
  {\bibfield  {journal} {\bibinfo  {journal} {Ann. Phys.}\ }\textbf {\bibinfo
  {volume} {318}},\ \bibinfo {pages} {316} (\bibinfo {year}
  {2005})}\BibitemShut {NoStop}%
\bibitem [{\citenamefont {Touchette}(2009)}]{Touchette2009}%
  \BibitemOpen
  \bibfield  {author} {\bibinfo {author} {\bibfnamefont {H.}~\bibnamefont
  {Touchette}},\ }\href {\doibase 10.1016/j.physrep.2009.05.002} {\bibfield
  {journal} {\bibinfo  {journal} {Phys. Rep.}\ }\textbf {\bibinfo {volume}
  {478}},\ \bibinfo {pages} {1} (\bibinfo {year} {2009})}\BibitemShut {NoStop}%
\bibitem [{\citenamefont {Lecomte}\ \emph {et~al.}(2007)\citenamefont
  {Lecomte}, \citenamefont {Appert-Rolland},\ and\ \citenamefont {van
  Wijland}}]{Lecomte2007}%
  \BibitemOpen
  \bibfield  {author} {\bibinfo {author} {\bibfnamefont {V.}~\bibnamefont
  {Lecomte}}, \bibinfo {author} {\bibfnamefont {C.}~\bibnamefont
  {Appert-Rolland}}, \ and\ \bibinfo {author} {\bibfnamefont {F.}~\bibnamefont
  {van Wijland}},\ }\href {\doibase DOI 10.1007/s10955-006-9254-0} {\bibfield
  {journal} {\bibinfo  {journal} {J. Stat. Phys.}\ }\textbf {\bibinfo {volume}
  {127}},\ \bibinfo {pages} {51} (\bibinfo {year} {2007})}\BibitemShut
  {NoStop}%
\bibitem [{\citenamefont {Garrahan}\ \emph {et~al.}(2007)\citenamefont
  {Garrahan}, \citenamefont {Jack}, \citenamefont {Lecomte}, \citenamefont
  {Pitard}, \citenamefont {van Duijvendijk},\ and\ \citenamefont {van
  Wijland}}]{Garrahan2007}%
  \BibitemOpen
  \bibfield  {author} {\bibinfo {author} {\bibfnamefont {J.~P.}\ \bibnamefont
  {Garrahan}}, \bibinfo {author} {\bibfnamefont {R.~L.}\ \bibnamefont {Jack}},
  \bibinfo {author} {\bibfnamefont {V.}~\bibnamefont {Lecomte}}, \bibinfo
  {author} {\bibfnamefont {E.}~\bibnamefont {Pitard}}, \bibinfo {author}
  {\bibfnamefont {K.}~\bibnamefont {van Duijvendijk}}, \ and\ \bibinfo {author}
  {\bibfnamefont {F.}~\bibnamefont {van Wijland}},\ }\href {\doibase
  10.1103/PhysRevLett.98.195702} {\bibfield  {journal} {\bibinfo  {journal}
  {Phys. Rev. Lett.}\ }\textbf {\bibinfo {volume} {98}},\ \bibinfo {pages}
  {195702} (\bibinfo {year} {2007})}\BibitemShut {NoStop}%
\bibitem [{\citenamefont {Garrahan}\ \emph {et~al.}(2009)\citenamefont
  {Garrahan}, \citenamefont {Jack}, \citenamefont {Lecomte}, \citenamefont
  {Pitard}, \citenamefont {van Duijvendijk},\ and\ \citenamefont {van
  Wijland}}]{Garrahan2009}%
  \BibitemOpen
  \bibfield  {author} {\bibinfo {author} {\bibfnamefont {J.~P.}\ \bibnamefont
  {Garrahan}}, \bibinfo {author} {\bibfnamefont {R.~L.}\ \bibnamefont {Jack}},
  \bibinfo {author} {\bibfnamefont {V.}~\bibnamefont {Lecomte}}, \bibinfo
  {author} {\bibfnamefont {E.}~\bibnamefont {Pitard}}, \bibinfo {author}
  {\bibfnamefont {K.}~\bibnamefont {van Duijvendijk}}, \ and\ \bibinfo {author}
  {\bibfnamefont {F.}~\bibnamefont {van Wijland}},\ }\href@noop {} {\bibfield
  {journal} {\bibinfo  {journal} {J. Phys. A}\ }\textbf {\bibinfo {volume}
  {42}},\ \bibinfo {pages} {075007} (\bibinfo {year} {2009})}\BibitemShut
  {NoStop}%
\bibitem [{\citenamefont {Garrahan}\ and\ \citenamefont
  {Chandler}(2002)}]{Garrahan2002}%
  \BibitemOpen
  \bibfield  {author} {\bibinfo {author} {\bibfnamefont {J.}~\bibnamefont
  {Garrahan}}\ and\ \bibinfo {author} {\bibfnamefont {D.}~\bibnamefont
  {Chandler}},\ }\href {\doibase 035704} {\bibfield  {journal} {\bibinfo
  {journal} {Phys. Rev. Lett.}\ }\textbf {\bibinfo {volume} {89}} (\bibinfo
  {year} {2002}),\ 035704}\BibitemShut {NoStop}%
\bibitem [{\citenamefont {Eisert}\ \emph {et~al.}(2014)\citenamefont {Eisert},
  \citenamefont {Friesdorf},\ and\ \citenamefont {Gogolin}}]{Eisert2014}%
  \BibitemOpen
  \bibfield  {author} {\bibinfo {author} {\bibfnamefont {J.}~\bibnamefont
  {Eisert}}, \bibinfo {author} {\bibfnamefont {M.}~\bibnamefont {Friesdorf}}, \
  and\ \bibinfo {author} {\bibfnamefont {C.}~\bibnamefont {Gogolin}},\ }\href
  {\doibase 10.1038/NPHYS3215} {\bibfield  {journal} {\bibinfo  {journal}
  {Nature Physics}\ }\textbf {\bibinfo {volume} {11}},\ \bibinfo {pages} {7}
  (\bibinfo {year} {2014})}\BibitemShut {NoStop}%
\bibitem [{\citenamefont {Gogolin}\ and\ \citenamefont
  {Eisert}()}]{Gogolin2015}%
  \BibitemOpen
  \bibfield  {author} {\bibinfo {author} {\bibfnamefont {C.}~\bibnamefont
  {Gogolin}}\ and\ \bibinfo {author} {\bibfnamefont {J.}~\bibnamefont
  {Eisert}},\ }\href@noop {} {\ }\Eprint {http://arxiv.org/abs/1503.07538}
  {arXiv:1503.07538} \BibitemShut {NoStop}%
\bibitem [{\citenamefont {Brown}\ \emph {et~al.}(2008)\citenamefont {Brown},
  \citenamefont {Santos}, \citenamefont {Starling},\ and\ \citenamefont
  {Viola}}]{Brown2008}%
  \BibitemOpen
  \bibfield  {author} {\bibinfo {author} {\bibfnamefont {W.~G.}\ \bibnamefont
  {Brown}}, \bibinfo {author} {\bibfnamefont {L.~F.}\ \bibnamefont {Santos}},
  \bibinfo {author} {\bibfnamefont {D.~J.}\ \bibnamefont {Starling}}, \ and\
  \bibinfo {author} {\bibfnamefont {L.}~\bibnamefont {Viola}},\ }\href
  {\doibase 10.1103/PhysRevE.77.021106} {\bibfield  {journal} {\bibinfo
  {journal} {Phys. Rev. E}\ }\textbf {\bibinfo {volume} {77}},\ \bibinfo
  {pages} {021106} (\bibinfo {year} {2008})}\BibitemShut {NoStop}%
\bibitem [{\citenamefont {Dukesz}\ \emph {et~al.}(2009)\citenamefont {Dukesz},
  \citenamefont {Zilbergerts},\ and\ \citenamefont {Santos}}]{Dukesz2009}%
  \BibitemOpen
  \bibfield  {author} {\bibinfo {author} {\bibfnamefont {F.}~\bibnamefont
  {Dukesz}}, \bibinfo {author} {\bibfnamefont {M.}~\bibnamefont {Zilbergerts}},
  \ and\ \bibinfo {author} {\bibfnamefont {L.~F.}\ \bibnamefont {Santos}},\
  }\href {http://stacks.iop.org/1367-2630/11/i=4/a=043026} {\bibfield
  {journal} {\bibinfo  {journal} {New J. Phys.}\ }\textbf {\bibinfo {volume}
  {11}},\ \bibinfo {pages} {043026} (\bibinfo {year} {2009})}\BibitemShut
  {NoStop}%
\bibitem [{\citenamefont {Haake}(2010)}]{Haake2010}%
  \BibitemOpen
  \bibfield  {author} {\bibinfo {author} {\bibfnamefont {F.}~\bibnamefont
  {Haake}},\ }\href {\doibase 10.1007/978-3-642-05428-0} {\emph {\bibinfo
  {title} {{Quantum Signatures of Chaos}}}},\ \bibinfo {series} {Springer
  Series in Synergetics}, Vol.~\bibinfo {volume} {54}\ (\bibinfo  {publisher}
  {Springer Berlin Heidelberg},\ \bibinfo {address} {Berlin, Heidelberg},\
  \bibinfo {year} {2010})\BibitemShut {NoStop}%
\bibitem [{Note1()}]{Note1}%
  \BibitemOpen
  \bibinfo {note} {Many-body quantum system are expected to ``equilibrate''
  \cite {Gogolin2015}, in the sense of the state becoming close to the
  time-averaged state, at long enough times under fairly general conditions
  (such as no energy gap degeneracy). This is independent on whether
  thermalisation is achieved or not (for example this so-called equilibrium
  state could be initial state dependent).}\BibitemShut {Stop}%
\bibitem [{\citenamefont {Oganesyan}\ and\ \citenamefont
  {Huse}(2007{\natexlab{b}})}]{Oganesyan2007}%
  \BibitemOpen
  \bibfield  {author} {\bibinfo {author} {\bibfnamefont {V.}~\bibnamefont
  {Oganesyan}}\ and\ \bibinfo {author} {\bibfnamefont {D.~A.}\ \bibnamefont
  {Huse}},\ }\href {\doibase 10.1103/PhysRevB.75.155111} {\bibfield  {journal}
  {\bibinfo  {journal} {Phys. Rev. B}\ }\textbf {\bibinfo {volume} {75}},\
  \bibinfo {pages} {155111} (\bibinfo {year} {2007}{\natexlab{b}})}\BibitemShut
  {NoStop}%
\bibitem [{\citenamefont {Gogolin}\ \emph {et~al.}(2011)\citenamefont
  {Gogolin}, \citenamefont {M\"uller},\ and\ \citenamefont
  {Eisert}}]{Gogolin2011}%
  \BibitemOpen
  \bibfield  {author} {\bibinfo {author} {\bibfnamefont {C.}~\bibnamefont
  {Gogolin}}, \bibinfo {author} {\bibfnamefont {M.~P.}\ \bibnamefont
  {M\"uller}}, \ and\ \bibinfo {author} {\bibfnamefont {J.}~\bibnamefont
  {Eisert}},\ }\href {\doibase 10.1103/PhysRevLett.106.040401} {\bibfield
  {journal} {\bibinfo  {journal} {Phys. Rev. Lett.}\ }\textbf {\bibinfo
  {volume} {106}},\ \bibinfo {pages} {040401} (\bibinfo {year}
  {2011})}\BibitemShut {NoStop}%
\bibitem [{\citenamefont {Rigol}\ \emph {et~al.}(2008)\citenamefont {Rigol},
  \citenamefont {Dunjko},\ and\ \citenamefont {Olshanii}}]{Rigol2008}%
  \BibitemOpen
  \bibfield  {author} {\bibinfo {author} {\bibfnamefont {M.}~\bibnamefont
  {Rigol}}, \bibinfo {author} {\bibfnamefont {V.}~\bibnamefont {Dunjko}}, \
  and\ \bibinfo {author} {\bibfnamefont {M.}~\bibnamefont {Olshanii}},\ }\href
  {\doibase 10.1038/nature06838} {\bibfield  {journal} {\bibinfo  {journal}
  {Nature}\ }\textbf {\bibinfo {volume} {452}},\ \bibinfo {pages} {854}
  (\bibinfo {year} {2008})}\BibitemShut {NoStop}%
\bibitem [{\citenamefont {Rigol}\ and\ \citenamefont
  {Srednicki}(2012)}]{Rigol2012}%
  \BibitemOpen
  \bibfield  {author} {\bibinfo {author} {\bibfnamefont {M.}~\bibnamefont
  {Rigol}}\ and\ \bibinfo {author} {\bibfnamefont {M.}~\bibnamefont
  {Srednicki}},\ }\href {\doibase 10.1103/PhysRevLett.108.110601} {\bibfield
  {journal} {\bibinfo  {journal} {Phys. Rev. Lett.}\ }\textbf {\bibinfo
  {volume} {108}},\ \bibinfo {pages} {110601} (\bibinfo {year}
  {2012})}\BibitemShut {NoStop}%
\bibitem [{\citenamefont {Srednicki}(1994)}]{Srednicki1994}%
  \BibitemOpen
  \bibfield  {author} {\bibinfo {author} {\bibfnamefont {M.}~\bibnamefont
  {Srednicki}},\ }\href {\doibase 10.1103/PhysRevE.50.888} {\bibfield
  {journal} {\bibinfo  {journal} {Phys. Rev. E}\ }\textbf {\bibinfo {volume}
  {50}},\ \bibinfo {pages} {888} (\bibinfo {year} {1994})}\BibitemShut
  {NoStop}%
\bibitem [{\citenamefont {Deutsch}(1991)}]{Deutsch1991}%
  \BibitemOpen
  \bibfield  {author} {\bibinfo {author} {\bibfnamefont {J.~M.}\ \bibnamefont
  {Deutsch}},\ }\href {\doibase 10.1103/PhysRevA.43.2046} {\bibfield  {journal}
  {\bibinfo  {journal} {Phys. Rev. A}\ }\textbf {\bibinfo {volume} {43}},\
  \bibinfo {pages} {2046} (\bibinfo {year} {1991})}\BibitemShut {NoStop}%
\bibitem [{\citenamefont {Biroli}\ \emph {et~al.}(2010)\citenamefont {Biroli},
  \citenamefont {Kollath},\ and\ \citenamefont {L\"auchli}}]{Biroli2010}%
  \BibitemOpen
  \bibfield  {author} {\bibinfo {author} {\bibfnamefont {G.}~\bibnamefont
  {Biroli}}, \bibinfo {author} {\bibfnamefont {C.}~\bibnamefont {Kollath}}, \
  and\ \bibinfo {author} {\bibfnamefont {A.~M.}\ \bibnamefont {L\"auchli}},\
  }\href {\doibase 10.1103/PhysRevLett.105.250401} {\bibfield  {journal}
  {\bibinfo  {journal} {Phys. Rev. Lett.}\ }\textbf {\bibinfo {volume} {105}},\
  \bibinfo {pages} {250401} (\bibinfo {year} {2010})}\BibitemShut {NoStop}%
\bibitem [{\citenamefont {Znidaric}\ \emph {et~al.}(2008)\citenamefont
  {Znidaric}, \citenamefont {Prosen},\ and\ \citenamefont
  {Prelovsek}}]{Znidaric2008}%
  \BibitemOpen
  \bibfield  {author} {\bibinfo {author} {\bibfnamefont {M.}~\bibnamefont
  {Znidaric}}, \bibinfo {author} {\bibfnamefont {T.}~\bibnamefont {Prosen}}, \
  and\ \bibinfo {author} {\bibfnamefont {P.}~\bibnamefont {Prelovsek}},\
  }\href@noop {} {\bibfield  {journal} {\bibinfo  {journal} {Phys. Rev. B}\
  }\textbf {\bibinfo {volume} {77}},\ \bibinfo {pages} {064426} (\bibinfo
  {year} {2008})}\BibitemShut {NoStop}%
\bibitem [{\citenamefont {Nanduri}\ \emph {et~al.}(2014)\citenamefont
  {Nanduri}, \citenamefont {Kim},\ and\ \citenamefont {Huse}}]{Nanduri2014}%
  \BibitemOpen
  \bibfield  {author} {\bibinfo {author} {\bibfnamefont {A.}~\bibnamefont
  {Nanduri}}, \bibinfo {author} {\bibfnamefont {H.}~\bibnamefont {Kim}}, \ and\
  \bibinfo {author} {\bibfnamefont {D.~A.}\ \bibnamefont {Huse}},\ }\href
  {\doibase 10.1103/PhysRevB.90.064201} {\bibfield  {journal} {\bibinfo
  {journal} {Phys. Rev. B}\ }\textbf {\bibinfo {volume} {90}},\ \bibinfo
  {pages} {064201} (\bibinfo {year} {2014})}\BibitemShut {NoStop}%
\bibitem [{\citenamefont {Kj\"all}\ \emph {et~al.}(2014)\citenamefont
  {Kj\"all}, \citenamefont {Bardarson},\ and\ \citenamefont
  {Pollmann}}]{Kjall2014}%
  \BibitemOpen
  \bibfield  {author} {\bibinfo {author} {\bibfnamefont {J.~A.}\ \bibnamefont
  {Kj\"all}}, \bibinfo {author} {\bibfnamefont {J.~H.}\ \bibnamefont
  {Bardarson}}, \ and\ \bibinfo {author} {\bibfnamefont {F.}~\bibnamefont
  {Pollmann}},\ }\href {\doibase 10.1103/PhysRevLett.113.107204} {\bibfield
  {journal} {\bibinfo  {journal} {Phys. Rev. Lett.}\ }\textbf {\bibinfo
  {volume} {113}},\ \bibinfo {pages} {107204} (\bibinfo {year}
  {2014})}\BibitemShut {NoStop}%
\bibitem [{\citenamefont {Vosk}\ and\ \citenamefont {Altman}(2014)}]{Vosk2014}%
  \BibitemOpen
  \bibfield  {author} {\bibinfo {author} {\bibfnamefont {R.}~\bibnamefont
  {Vosk}}\ and\ \bibinfo {author} {\bibfnamefont {E.}~\bibnamefont {Altman}},\
  }\href {\doibase 10.1103/PhysRevLett.112.217204} {\bibfield  {journal}
  {\bibinfo  {journal} {Phys. Rev. Lett.}\ }\textbf {\bibinfo {volume} {112}},\
  \bibinfo {pages} {217204} (\bibinfo {year} {2014})}\BibitemShut {NoStop}%
\bibitem [{\citenamefont {Friesdorf}\ \emph {et~al.}(2014)\citenamefont
  {Friesdorf}, \citenamefont {Werner}, \citenamefont {Goihl}, \citenamefont
  {Eisert},\ and\ \citenamefont {Brown}}]{Friesdorf2014}%
  \BibitemOpen
  \bibfield  {author} {\bibinfo {author} {\bibfnamefont {M.}~\bibnamefont
  {Friesdorf}}, \bibinfo {author} {\bibfnamefont {A.~H.}\ \bibnamefont
  {Werner}}, \bibinfo {author} {\bibfnamefont {M.}~\bibnamefont {Goihl}},
  \bibinfo {author} {\bibfnamefont {J.}~\bibnamefont {Eisert}}, \ and\ \bibinfo
  {author} {\bibfnamefont {W.}~\bibnamefont {Brown}},\ }\href
  {http://arxiv.org/abs/1412.5605} {\  (\bibinfo {year} {2014})},\ \Eprint
  {http://arxiv.org/abs/1412.5605} {arXiv:1412.5605} \BibitemShut {NoStop}%
\bibitem [{Note2()}]{Note2}%
  \BibitemOpen
  \bibinfo {note} {For small values of $N$ we define $\protect \overline
  {S}_{\protect \text {eq}}$ as the time average over the final plateau,
  ensuring that the system does not undergo renewals.}\BibitemShut {Stop}%
\end{thebibliography}
\end{document}